# Chiral selection rules for multi-photon processes in two-dimensional honeycomb materials

Jingxin Cheng,[1] Di Huang,[1] Tao Jiang,[1] Yuwei Shan,[1] Yingguo Li,[1] Shiwei Wu,[1,2,*] Wei-Tao Liu[1,2,*]

[1]State Key Laboratory of Surface Physics, Key Laboratory of Micro and Nano Photonic Structures (MOE), and Department of Physics, Fudan University, Shanghai 200433, China

[2]Collaborative Innovation Center of Advanced Microstructures, Nanjing 210093, China

*Corresponding author: wtliu@fudan.edu.cn, swwu@fudan.edu.cn

**Abstract:**

We examined the chirality dependent optical selection rules in two-dimensional monolayer materials with the honeycomb lattice, and based on symmetry argument, we generalized these rules to multi-photon transitions of arbitrary orders. We also presented the phase relations between incident and outgoing photons in such processes. The results agreed nicely with our experimental observations of second and third harmonic generations. In particular, we demonstrated that the phase relation of chiral second harmonic generation can serve as a handy tool for imaging domains and domain boundaries of these monolayers. Our results can benefit future studies on chirality related optical phenomena and opto-electronic applications of such materials.

Two dimensional (2D) materials with honeycomb lattices are of great interest in recent years for their peculiar optical and electronic properties [1-3]. They include graphene, hexagonal boron nitride (h-BN), transitional metal dichalcogenides (TMDCs, including $MoS_2$, $MoSe_2$, $WS_2$, $WSe_2$), and monochalcogenides (including GaSe and InSe). All these monolayers share similar honeycomb structure like graphene as viewed from the top [Fig. 1(a)], which generates the opposite Berry curvature and orbital magnetic moment at *K* and *K'* valleys [4-8]. In TMDCs, the two sub-lattices are occupied alternatively by different atoms, which breaks the inversion symmetry and, together with the spin-orbital coupling, leads to the well-known valley contrasting optical selection rule [9-11]. As illustrated in Fig. 1(b), the one-photon excitations at the *K* and *K'* points must be driven by the left and right circularly polarized (CP) photons (denoted by $\sigma_+$ and $\sigma_-$), respectively. This can be evident from the chirality resolved photoluminescence (PL) spectra from the TMDC monolayer, such as $WSe_2$ [Fig. 1(c)]. The excitation wavelength and power were 1.88 eV and 10 μW from a 150 fs Newport Inspire auto 100 OPO laser system [12, 13]. At 19 K, upon the $\sigma_+$ excitation, the PL from the valley is also dominated by the $\sigma_+$ component [9-11]. By defining the chirality parameter to be [10, 11]

$$\rho = \frac{I(\sigma_+) - I(\sigma_-)}{I(\sigma_+) + I(\sigma_-)}$$

with *I* being the intensity, we found $\rho \approx 0.3$ for the PL, which quantifies the preservation of valley polarization.

Later, several groups observed chirality dependent second harmonic generation (SHG) and two-photon photoluminescence (TPPL) in these materials, and attributed the phenomena to the conservation of both valley and orbital angular momenta [14-16]. Here we revisit them from the viewpoint of the group theory. For all such honeycomb lattices, the local symmetry at *K* and *K'* valleys belong to the $C_{3h}$ double group with a three-fold rotational symmetry [17]. Upon incidence, the electric field vectors of a $\sigma_+$ ($\sigma_-$) photon can be described by $\hat{x} \pm i\hat{y}$, with $\hat{x}, \hat{y}$ being unit vectors in the sample plane [Fig. 2(a)]. Following the notations in Ref. [18], the $\sigma_+$ and $\sigma_-$ light fields transform according to the irreducible representations $\Gamma_2$ and $\Gamma_3$, respectively. From the multiplication table for the $C_{3h}$ group, we find $\Gamma_2 \otimes \Gamma_2 = \Gamma_3$, and $\Gamma_3 \otimes \Gamma_3 = \Gamma_2$. This means the sequential transition of two $\sigma_+$ photons would obey the same selection rule with that for one $\sigma_-$ photon, and vice versa, which sets the foundation of the chiral selection rule for 2D honeycomb lattices.

By applying the above rule to the two-photon process, we expect two $\sigma_-$ ($\sigma_+$) photons are needed to excite the transition near the *K* (*K'*) point. For the second harmonic generation (SHG), it is followed by the emission of one $\sigma_+$ ($\sigma_-$) photon at the SH frequency [Fig. 2(b)]. This is indeed the case as shown in Fig. 2(c): with the $\sigma_+$ input, we observed the SHG to be predominantly $\sigma_-$ polarized (left panel); and with $\sigma_-$ input, the SHG was predominantly $\sigma_+$ polarized (right panel). Remarkably, the chirality parameter is nearly 1.0 even at the room temperature [15, 16], which far exceeds that of the one-photon PL. This is because the PL lifetime is that of the non-equilibrium carrier population; during the emission, it is vulnerable to the inter-valley scattering that kills the valley polarization. Instead, the SHG is an instantaneous process, only lasts until the excited carriers loses coherence. Therefore, compared to the PL, the CP-SHG can provide the information of a nearly perfect valley polarization.

The CP-SHG is also distinct from the usual LP-SHG in its anisotropy feature. In the lab coordinates, we define $\phi$ to be the angle between the $\hat{y}$-axis of the incident light field and $\hat{y}'$-axis, the armchair axis (along the mirror plane) of the honeycomb lattice [Fig.

2(d)]. $\phi$ therefore represents the phase of the incident CP light upon hitting the sample plane. The reflection operation against the y'-z' plane (z' || z being the sample surface normal) transforms a $\sigma_+$ photon into a $\sigma_-$ photon with $\phi \to -\phi$. Since the y'-z' is a mirror plane of the lattice, the sample is invariant after this operation, so all the physics must be the same for a $\sigma_+$ photon of the phase $\phi$, and for a $\sigma_-$ photon of the phase $-\phi$. The SHG accumulates the phases of two incident photons ($\phi + \phi = 2\phi$); and because of the opposite chirality, the emitted SH photon will have a phase of $-2\phi$ [Fig. 2(d)]. So for a uniform honeycomb monolayer, if we rotate it azimuthally with respect to the surface normal, the SH output changes only the phase, but not the intensity, as seen in Fig. 2(e). This is in sharp contrast to the LP-SHG responses, which exhibit 6-fold anisotropic patterns with respect to $\phi$ [Fig. 2(f)] [12, 13, 19-23].

On the other hand, if the monolayer consists of multiple domains with different orientations, the SHG from neighboring domains will have different phases and interfere. If the armchair directions of neighboring domains are off by an angle $\Delta\theta$, the phase difference between their SH signal is then $\Delta\phi = 3\Delta\theta$, and the SHG intensity at the domain boundary is

$$I_b = I_0 \cdot \cos(\tfrac{3}{2}\Delta\theta)^2, \tag{1}$$

with $I_0$ being the SH intensity of a single domain. So domain boundaries will always appear dimmer than neighboring domains. Figure 3(a) shows the white-light microscopy image of a CVD-grown multi-domain $MoS_2$ monolayer, on which no grain boundary could be resolved. Figure 3(b) is the LP-SHG image of the same area. As we reported in [13], under a fixed polarization combination, domains of different orientations show different contrasts and exposing the domain boundaries. Yet because of the anisotropic response, some domains and domain boundaries are hardly visible. With CP-SHG [Fig. 3(c)], now all domains are simultaneously visible at nearly the same intensity, and all domain boundaries can be seen due to the interference discussed above. From the SH intensity profile along line-cuts across multiple domains [for example, the blue line in Fig. 3(d)], we can obtain $I_b$ for each domain boundary as a function of the $\Delta\theta$ between its neighboring domains, as plotted in [Fig. 3(e)]. Here $\Delta\theta$ is taken between 0 and $60°$ due to the 3-fold symmetry of honeycomb lattice, and can be obtained from either the LP-SHG anisotropy pattern, or the intersection angle between neighboring domains [13]. The relation $I_b(\Delta\theta)$ can be fitted very nicely by Eq. (1) [solid curve in Fig. 3(e)].

Now we discuss the case for three-photon excitations, such as the excitation of THG. For the $C_{3h}$ group, we have $\Gamma_2\otimes\Gamma_3=\Gamma_3\otimes\Gamma_2=\Gamma_1$, with $\Gamma_1$ being the totally symmetric representation. So if the excitation is purely $\sigma_+$ polarized, the three-photon excitation transforms according to $\Gamma_2\otimes\Gamma_2\otimes\Gamma_2=\Gamma_3\otimes\Gamma_2=\Gamma_1$. As it does not contain either $\Gamma_2$ or $\Gamma_3$ component, it cannot lead to a direct transition at either $K$ or $K'$ valley [Fig. 4(a)]. The same applies to $\sigma_-$ photons [Fig. 4(a)]. Indeed as seen in Fig. 4(b), the THG signal is nearly zero upon CP excitations (blue curve).

In contrast, strong THG emerges upon the LP excitation [Fig. 4(b), red curve]. This is because the LP input mixes both $\sigma_+$ and $\sigma_-$ photons; as $\Gamma_2\otimes\Gamma_2\otimes\Gamma_3=\Gamma_2$, the transition by sequentially absorbing two $\sigma_+$ photons and one $\sigma_-$ photon is allowed at the $K$ valley, and lead to the emission of a $\sigma_+$ photon [Fig. 4(c), left panel]. Similarly, because of $\Gamma_3\otimes\Gamma_3\otimes\Gamma_2=\Gamma_3$, the transition with two $\sigma_-$ photons and one $\sigma_+$ photon is allowed at the $K'$ valley, and lead to the emission of a $\sigma_-$ photon [Fig. 4(c), right panel]. In total there are three pairs of $\sigma_+$ and $\sigma_-$ photons involved during the excitation, and one pair of $\sigma_+$ and $\sigma_-$ photons for the emission. Therefore, the THG signal upon LP excitation is also LP, which is evident from the polarization patterns in Fig. 4(d). Meanwhile, for the excitation at the $K$ valley, if the incident $\sigma_+$ photon has a phase $\phi$, the phase of the emitted $\sigma_+$ THG photon is then $2\phi - \phi = \phi$ [Fig. 4(e), left panel]. Similarly, the $\sigma_-$ THG photon emitted from the $K$' valley has a phase of $-\phi$ [Fig. 4(e), left panel]. The total LP-THG emission thus has the same phase with that of the input LP light, also agreeing well with our observation [Fig. 4(d)].

As we mentioned above, $\Gamma_1$ is the totally symmetric representation, so its product with any other representation leaves that representation unchanged. Therefore, the above chiral selection rules can be easily generalized to direct transitions with more than three photons. Below we summarize the rules for multi-photon transitions in a 2D honeycomb lattice around valleys: 1) the ($3m$-2)-photon ($m \in$ positive integers) transition can be excited by CP photons, and the emission is of the opposite chirality; 2) the ($3m$-1)-photon transition can also be excited by CP photons, but the direct emission is a photon of the same chirality; 3) the ($3m$)-photon transition cannot be excited by CP photons of the same chirality.

To conclude, we studied the chiral selection rules for multi-photon transitions in 2D honeycomb lattices, which are directly consequent from the local symmetry of $K$/$K$' valleys of such lattices. We also presented the phase relation between input and output photons in different cases. In particular, the phase relation of CP-SHG allows a nice tool for simultaneously imaging all TMDC domains and domain boundaries. More generally, these chiral selection rules are robust as protected by symmetry, so it can be readily generalized to high order direct transitions with more photons. For example, it can help guide the generation of CP high harmonics using related materials [24-26].

**Funding.** The work at Fudan University was supported by the National Basic Research Program of China (Grant No. 2014CB921601), National Key Research and Development Program of China (Grant Nos. 2016YFA0300900, 2016YFA0301002), National Natural Science Foundation of China (Grant Nos. 11622429). W.-T. L. also acknowledges support from the National Program for Support of Top-Notch Young Professionals and "Shuguang Program" supported by Shanghai Education Development Foundation and Shanghai Municipal Education Commission.

**Acknowledgment**. We thank Professor Yuen-Ron Shen for insightful discussion, and Professor Yanfeng Zhang for providing the CVD grown $MoS_2$ monolayer samples.

**Figures and Captions**

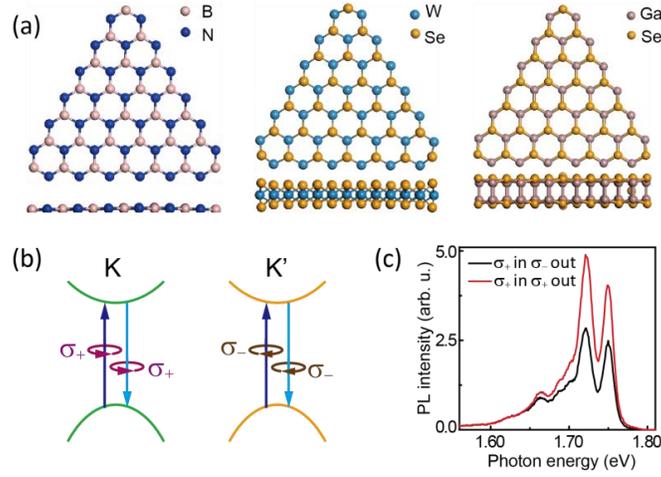

**Fig. 1.** (a) Left to right: top and side views of the atomic structure of monolayer h-BN, WSe$_2$ and GaSe. (b) Valley contrasting selection rules for one photon transition in 2D honeycomb materials. At the $K$ ($K'$) valley, only $\sigma_+$ ($\sigma_-$) light can be absorbed and emitted. (c) Chirality resolved PL spectrum from a monolayer WSe$_2$ at 19 K. Red and black curves are the PL spectra with $\sigma_+$ and $\sigma_-$ chirality upon the $\sigma_+$ excitation.

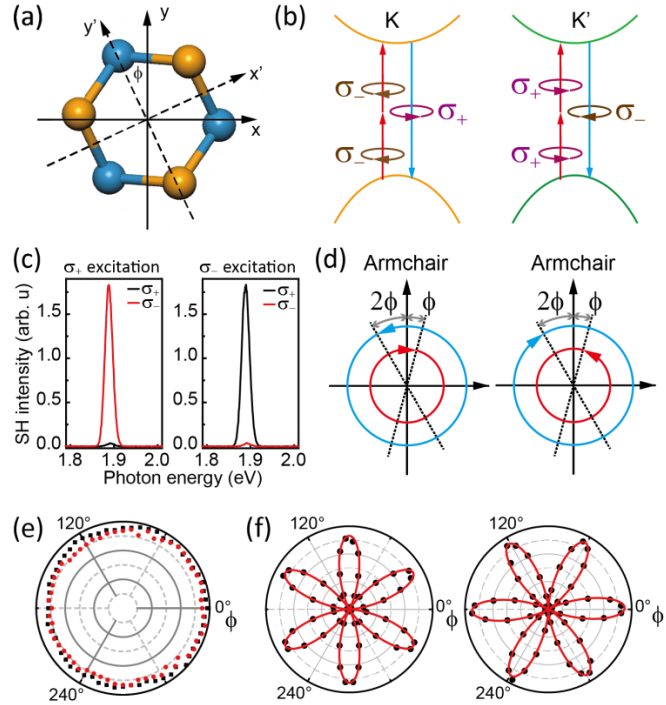

**Fig. 2.** (a) Definition of coordinates. The optical field of input CP light is along *y* upon hitting the sample plane, and *y'* is the armchair direction of the crystal and *x'* is zig-zag direction. $\phi$ is the angle between *y* and *y'*, referred to as the phase of the CP light. (b) SHG selection rules in 2D honeycomb lattice. Absorption of two $\sigma_-$ ($\sigma_+$) photons at *K* (*K'*) valley lead to the emission of one $\sigma_+$ ($\sigma_-$) photon at twice the incoming photon energy. (c) Chirality resolved SHG spectra from monolayer $WSe_2$ with circularly polarized fundamental beam at 0.94 eV and 0.14 mW. Red (black) curves are the $\sigma_-$ ($\sigma_+$) polarized SHG signal. (d) The chirality and relative phase between fundamental and SH photons. The phase of the emitted $\sigma_+$ ($\sigma_-$) photon is delayed (advanced) by $3\phi$ relative to the fundamental photon. (e) The CP-SHG intensity with $\sigma_+$ (red dots) or $\sigma_-$ (black dots) excitation versus the angle $\phi$. (f) The LP-SHG intensity versus the angle $\phi$, with the SH signal polarized perpendicularly (left) or parallel (right) to that of the excitation.

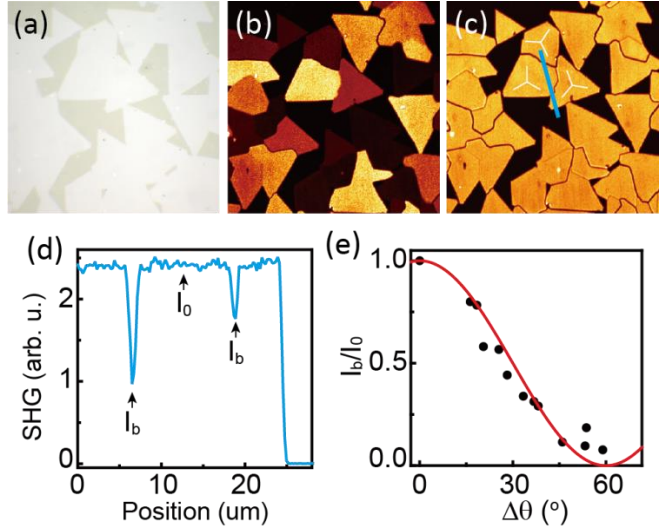

**Fig. 3.** (a) White-light microscopy image of CVD-grown MoS$_2$ monolayer on a sapphire substrate. (b) LP-SHG microscopy images of the same area in (a). (c) CP-SHG microscopy images of the same area in (a). (d) Profile of the SHG intensity along the blue line-cut in (c). $I_0$ denotes the intensity inside a single domain, $I_b$ denote those at domain boundaries. (e) Black dots: the CP-SHG intensity of domain boundaries (normalized to that of a single domain), versus the offset angle between armchair orientations [marked by white lines in (c)] of neighboring domains. The red curve is a fit using Eq. (1).

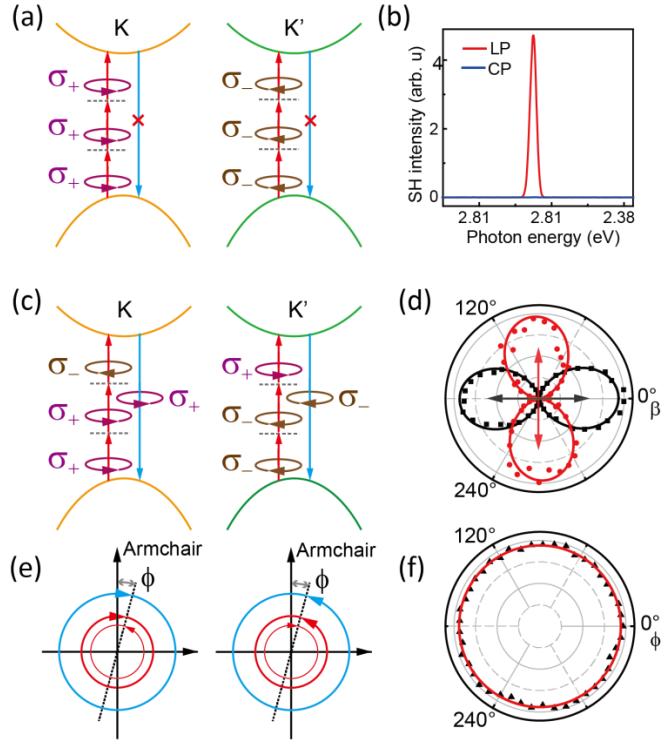

**Fig. 4.** (a) THG process in 2D honeycomb lattice with purely CP excitations. (b) THG spectra under CP (dark blue curves) and LP (red curve) excitations from monolayer $WS_2$ at the room temperature. The excitation is at 0.89 eV and 0.22 mW. (c) THG process under LP excitations. An LP excitation is the superposition of a $\sigma_+$ and $\sigma_-$ excitations (d) The polarization analysis of the LP-THG signal. The black and red symbols are THG intensity versus the analyzer angle $\beta$ with excitation beam polarized along different directions, as indicated by corresponding arrows. Solid curves are patterns expected for LP signals. (e) The chirality and relative phase between fundamental and TH photons. The phase of the emitted $\sigma_+$ ($\sigma_-$) photon is the same with that of the fundamental photon. (f) The LP-THG intensity versus the angle $\phi$.